\documentstyle[a4,12pt,amssymb]{article}
\begin{document}

\def\NPB#1#2#3{{\it Nucl.~Phys.} {\bf{B#1}} (19#2) #3}
\def\PLB#1#2#3{{\it Phys.~Lett.} {\bf{B#1}} (19#2) #3}
\def\PRD#1#2#3{{\it Phys.~Rev.} {\bf{D#1}} (19#2) #3}
\def\PRL#1#2#3{{\it Phys.~Rev.~Lett.} {\bf{#1}} (19#2) #3}
\def\ZPC#1#2#3{{\it Z.~Phys.} {\bf C#1} (19#2) #3}
\def\PTP#1#2#3{{\it Prog.~Theor.~Phys.} {\bf#1}  (19#2) #3}
\def\MPLA#1#2#3{{\it Mod.~Phys.~Lett.} {\bf#1} (19#2) #3}
\def\PR#1#2#3{{\it Phys.~Rep.} {\bf#1} (19#2) #3}
\def\AP#1#2#3{{\it Ann.~Phys.} {\bf#1} (19#2) #3}
\def\RMP#1#2#3{{\it Rev.~Mod.~Phys.} {\bf#1} (19#2) #3}
\def\HPA#1#2#3{{\it Helv.~Phys.~Acta} {\bf#1} (19#2) #3}
\def\JETPL#1#2#3{{\it JETP~Lett.} {\bf#1} (19#2) #3}
\def\JHEP#1#2#3{{\it J. High Energy Phys.} {\bf#1} (19#2) #3}

\def\reflist{\section*{References\markboth
        {REFLIST}{REFLIST}}\list
        {[\arabic{enumi}]\hfill}{\settowidth\labelwidth{[999]}
        \leftmargin\labelwidth
        \advance\leftmargin\labelsep\usecounter{enumi}}}
\let\endreflist\endlist \relax

\newcommand{\be}{\begin{equation}}
\newcommand{\ee}{\end{equation}}
\newcommand{\ba}{\begin{eqnarray}}
\newcommand{\ea}{\end{eqnarray}}
\newcommand{\dal}{\raisebox{0.085cm} 
{\fbox{\rule{0cm}{0.07cm}\,}}}
\newcommand{\dslash}{{\not\!\partial}}
\catcode`\@=11
\titlepage
\begin{flushright}
LPT-ORSAY 00/128 \\
hep-th/0012071
\end{flushright}
\vskip 1cm
\begin{center}
{\Huge \bf Consistent gravitino couplings in non-supersymmetric 
strings }
\end{center}
\vskip 1cm
\begin{center}

E. Dudas  and J. Mourad
\end{center} 
\vskip 0.5cm
\begin{center}
{\it Laboratoire de Physique Th\'eorique 
\footnote{Unit\'e Mixte de Recherche du CNRS (UMR 8627).},\\
 Universit\'e de Paris-Sud, B\^at. 210, F-91405 Orsay Cedex,
 France}
\end{center}
\vskip 2cm
\begin{center}
{\large Abstract}
\end{center}
The massless spectrum of the ten dimensional USp(32) 
type I string has an N=1 supergravity multiplet coupled 
to non-supersymmetric matter. This raises the question of the
consistency of the gravitino interactions.
We resolve this apparent puzzle by arguing
for the existence of a local supersymmetry which is non-linearly 
realised in the open  sector. We determine the 
leading order low energy effective Lagrangian.     
Similar results are expected to apply to lower dimensional type I models 
where supergravity is coupled to non-supersymmetric branes.
\noindent

\newpage
\section{Introduction and summary of results }

It is believed that in a non-supersymmetric string theory a massless spin 
$3/2$ gravitino cannot have consistent interactions \cite{gsw}. 
Indeed, local supersymmetry guarantees the consistency
of the gravitino equation of motion, as general covariance
does for the Einstein equations.
Consider the  gravitino field equation
\be
\Gamma^{MNP} D_N \Psi_P = J^M \ , \label{i1}
\ee
where $J_M$ is the source term for the gravitino. It implies,  
by taking a covariant derivative on both sides
of (\ref{i1}),
\be
D_M J^M = {1 \over 2} T_M^P \Gamma^M \Psi_P \ , \label{i2}
\ee 
where $T_M^P$ is the energy-momentum tensor, the source term in
the Einstein equations\footnote{We have neglected, for simplicity,
possible torsion contributions to (\ref{i2}).}. 
Eq. (\ref{i2}) is a strong constraint and it implies
that $J^M$ receives contributions from all matter fields.
A consistent way of insuring that (\ref{i2}) is fulfilled is 
to obtain it as a conservation law of some symmetry. 
It is in this
way that local supersymmetry provides consistent gravitino
interactions. The converse, i.e. that consistent gravitino
interactions imply supergravity, is very probable, although not proved.  

Recently, new Type I models were constructed
\cite{sugimoto,ads}, whose spectrum comprises a massless spin $3/2$
particle and non-supersymmetric matter. They are characterized by
tree-level supersymmetry in the closed string
spectrum and broken supersymmetry in the open string sector. 
The simplest model in this class is the 10d $USp(32)$ non-tachyonic
Type I string \cite{sugimoto}. It is constructed as 
an orientifold of type IIB with the group element
$\Omega'=\Omega \ (-1)^{F_{spacetime}}$ breaking half of the
supersymmetries in the closed (gravitational) sector, while 
completely breaking supersymmetry in the open (Yang-Mills) sector.
From the string point of view, 
the model is nevertheless consistent because it has standard spacetime
particle interpretation and it is possible to
cancel the Ramond-Ramond tadpole by choosing the gauge group $USp(32)$.
  
Lower dimensional Type I orbifold models were also constructed \cite{ads}.
Their closed spectrum is supersymmetric, but different from the
standard Type I compactifications. In addition, consistency rules ask for  
non-supersymmetric open spectra and result in tree-level scalar
potentials localized on the (anti)branes. 
 
At first sight, the simultaneous presence of a massless gravitino and
non-supersymmetric matter makes the consistency of the effective field 
theory difficult to achieve. Indeed, there is no
parameter to vary  that would restore supersymmetry in the limit when it goes
to zero\footnote{Strictly speaking, this is not quite true in
lower dimensional models, where by moving D-branes off the orientifold
fixed points, supersymmetry is restored at the massless level. However, in
the 10d model we discuss in detail here this is not the
case. Moreover, this (Wilson line) parameter
is not an ordinary order parameter. For a nonvanishing value
the massless D-brane spectra are supersymmetric, whereas for zero value
it becomes non-supersymmetric. This behaviour is opposite to that of an order
parameter.}. On the other hand, whereas in lower dimensional models 
the gravitino will
presumably acquire a mass through radiative corrections, in the 10d
$USp(32)$ case this cannot happen due to the Majorana-Weyl nature of the 10d
Type I gravitino. Indeed, a massive gravitino in 10d has 128 degrees of 
freedom, much more than the total fermionic, singlet under the
gauge group, degrees of freedom in this model.
The gravitino staying massless renders the consistency
of this model even more puzzling. 
 A first, complementary step undertaken recently was the determination
of the vacuum of the theory \cite{dm} and the study of its classical
stability \cite{dm3}, in the spirit of the Fishler-Susskind mechanism 
\cite{fs}.
    
The purpose of this paper is to prove the consistency of
the effective field theory of these models, by working out the low-energy
lagrangian for the Type I 10d model
\cite{sugimoto}. We argue that there is an underlying supersymmetry,
linearly realized in the closed sector and nonlinearly realized \`a la
Volkov-Akulov \cite{va} in the open $USp(32)$ sector, with the nonlinear
parameter being the string scale. In a loose sense, the corresponding
open spectra can be described as having supersymmetry broken at the
string scale. This distinguishes them from pure non-supersymmetric
strings \cite{dh}. 

A hint in favor of brane nonlinear supersymmetry is the existence of
a gauge singlet, called $\theta$ in the following, in the massless open 
spectrum. This fermion has the chirality of the gravitino, which makes 
it a good candidate for a goldstino of the non-linearly realized
local supersymmetry. Indeed, in the 10d $USp(32)$ model we
work out in detail below, the charged fermions belong to the antisymmetric
reducible representation containing a singlet. The presence of this 
goldstino allows the explicit construction of
a consistent effective Lagrangian describing the coupling of the $N=1$
supergravity to the dilaton tadpole and non-supersymmetric Yang-Mills 
matter. 

The 10d $USp(32)$ Type I model contains the gravitational multiplet 
with bosonic fields $(g_{MN}, B_{MN}, \phi)$ and fermions $(\Psi_M,
\lambda)$ in the closed spectrum. The open
spectrum contains gauge fields $A_M$ in the adjoint of the gauge
group, fermions $\chi$ in the ${\bf 495}$ and the fermion gauge singlet
$\theta$. 

Up to terms quadratic in fermions, the effective lagrangian of the
model in the Type I string metric, to be found explicitly in the following
sections, reads
\ba
L &\!\!=\!\!& L_{SUGRA} \!+\!  \sqrt{-G}
 e^{-\hat \phi} \Big[-{{1}\over{4}}G^{MP}G^{NQ} \ tr (F_{MN}F_{PQ}) 
 \!-\! \Lambda\Big] + \nonumber \\
&\!\!\! \sqrt{-g} \!\!\!\!& \Biggl\{ \!-\! {{1}\over{2}} tr \bar \chi
 \Gamma^M D_M \chi  \!+\! {{1}\over{16}} tr \bar \chi
\Gamma^{MNP}\chi H_{MNP} \!+\! T^{MNPQ} \ tr(F_{[MN}F_{PQ]}) \Biggr\}  \ , \label{i5}
\ea
where $G_{MN}$ and $\hat \phi$ are the modified Volkov-Akulov metric
and the dilaton, respectively, defined in (\ref{mod1}), (\ref{mod2})  
and we introduced the tensor notation
\ba
&&\!\!\!\!T^{MNPQ} \!=\!  -{1 \over 16} e^{-\phi} {\bar \theta} 
\Gamma^{MNPQR} (-\Psi_R \!+\! {1
\over 2} D_R \theta \!+\! {\sqrt{2} \over 6} \Gamma_R \lambda ) 
\nonumber \\ 
&\!\!-\!\!&\!\!{{\bar \theta} \over 16} \bigl[ 
({{1}\over{16}}  \Gamma^{MNPQN_1N_2N_3}
\!+\!\!{{9}\over{4}} g^{[MN_1}g^{NN_2}\Gamma^{PQN_3]}) H_{N_1N_2N_3}
\!\!+\! e^{-\phi} \Gamma^{[NPQ}\partial^{M]}\phi \bigr] \theta \ . \label{i6} 
\ea 
In (\ref{i5}), $L_{SUGRA}$ is the 10d $N=1$ supergravity
lagrangian (\ref{sugra}) and $H_{MNP}$ is the field strength of the
2-form $B_{MN}$, as defined in (\ref{om}). The lagrangian (\ref{i5}) is 
invariant under the supersymmetry transformations (\ref{linear}) 
(modified as in (\ref{extra})) and (\ref{volkov}). A more explicit form
of the Lagrangian (\ref{i5}) is displayed in (\ref{cs3}). Here we just
mention that the modified metric and dilaton render the Yang-Mills and
the tadpole terms in (\ref{i5}) supersymmetric and  the last term in
(\ref{i5}) is imposed by the Green-Schwarz anomaly cancellation mechanism. 

The main features can be simply understood by setting to zero all open
string fields except the fermion singlets and the three-form.  
The presence of the dilaton tadpole breaks supersymmetry 
and the lagrangian (\ref{i5}) reduces in this case to
\be
L = L_{SUGRA}-\Lambda\sqrt{-g} \ e^{-\phi} (1+ {1 \over 4} {\bar \theta} \Gamma^M D_M
 \theta - {1 \over 2}  {\bar \theta} \Gamma^M \Psi_M +
{1 \over \sqrt{2}} {\bar \theta} \lambda ) \ , \label{i3}
\ee
where $\Lambda = 64 T_9$ and $T_9$ is the D9 brane tension. 
It is transparent in  (\ref{i3}) how nonlinear supersymmetry or, 
equivalently, consistency of gravitino field equation, relates the
dilaton tadpole to goldstino - gravitino and goldstino - dilatino couplings.
Notice that the positive sign of the dilaton tadpole is required
in order to have a correct kinetic term for the goldstino $\theta$.
As disk NS-NS tadpoles are the main feature of models with supersymmetry
broken on branes, we expect a similar description in the lower-dimensional
models \cite{ads}.

The rest of the paper is devoted to the derivation of the
Lagrangian (\ref{i5}) and to exhibit a nice geometrical setting
underlying its construction. In Section 2,
we present the Volkov-Akulov construction 
of Lagrangians with non-linear global supersymmetry 
in a way which will be useful for its local generalisation.
In Section 3, we study the coupling of $N=1$ supergravity 
to the dilaton tadpole and the non-supersymmetric 
Yang-Mills field. We first determine the $\theta$ dependence
of the Lagrangian by requiring invariance under non-linear
local supersymmetry and then show that it can be obtained 
from a simple geometrical construction.
Contrary to the linear case, the Yang-Mills coupling to
supergravity does not require the modification of the Bianchi
identity of the three-form. On the other hand, anomaly
cancellation does require this modification.
In Section 4 we show that this modification is compatible with
non-linear supersymmetry and find the corresponding 
supersymmetric completion of the Lagrangian.  
In Section 5 
we provide another check on the existence of nonlinear
supersymmetry in the $USp(32)$ string by comparing
the results of the supersymmetric completion with a string theory
calculation of a three point amplitude.   
The generalization of the results of this paper to four dimensional 
models in the same class \cite{ads} is in work \cite{carlo}. 

Some related work in the literature concerns 
the description of partial supersymmetry breaking for BPS branes 
\cite{hlp}, the
nonlinear realization of supersymmetry for D9 branes \cite{abkz, brjo}
and also the world-volume actions for non-BPS branes in Type II 
strings \cite{sen}.

A possible application of the results of this paper is to 
brane-world scenarios \cite{lykken}. It would be interesting to
study details of models with tree-level (linearly realized)
supersymmetry in the bulk (gravitational) sector and non-linearly
realized supersymmetry on our brane universe\footnote{We would like to
thank T. Gherghetta for discussions on this issue.}. 

\section{Non-linear realisation of supersymmetry}

Let $\theta$ be a Weyl-Majorana spin 1/2 fermion in
10D  Minkowski space-time with the transformation rule
\be
\delta\theta=\eta+{{1}\over{4}} \ \bar\eta\Gamma^{M}\theta \ \partial_M
\theta,\label{tra}
\ee
where $\eta$ is a Majorana-Weyl constant spinor,
then the algebra is that of supersymmetry
\be
[\delta_{\xi_1},\delta_{\xi_2}]={{1}\over{2}}
\bar\xi_2\Gamma^M\xi_1
\partial_M \ .
\ee
Furthermore, the 1-forms defined by
\be
{\cal E}^{M}=dx^M+{{1}\over{4}}\bar\theta\Gamma^{M}d\theta,
\ee
transform, under (\ref{tra}), as
\be
\delta{\cal E}^M=L_{v}{\cal E}^{M} \ ,
\ee
where $L_v=di_v + i_v d$ ($i_v$ is the interior product with respect to
$v$) is the Lie derivative with respect to the vector field $v$ given by
\be 
v=-{{1}\over{4}}\bar \theta\Gamma^M\eta \ \partial_M \ . \label{lie}
\ee
The transformation of ${\cal E}^M$
has the same form as that of a moving basis under
a coordinate transformation and furthermore a Lorentz
rotation has the same effect on the moving basis as on ${\cal
E}^M$. So  the  lowest order lagrangian is given by
a cosmological constant-like term:
\be 
{\cal L}={\rm det} {\cal E}.
\ee
With this analogy in mind it is straightforward to add 
matter contributions to the action. The transformation rule of a
p-form field or its components is given by
\be
\delta A=L_v A \ , \ \delta A_{M_1 \cdots M_p} = ( \partial_{[M_1}v^N)
A_{NM_2 \cdots M_p]} + v^N \partial_N A_{M_1 \cdots M_p} \label{tran} 
\ee
and the lagrangian is obtained from the Poincar\'e
invariant lagrangian by replacing the Minkowski metric
by 
\be
{\cal G}_{MN}={\cal E}^a_M{\cal E}^b_N \eta_{ab} \ .
\ee
There is therefore a simple procedure 
to construct an action with non-linear supersymmetry
from a Poincar\'e invariant one: introduce the fermion 
$\theta$, the goldstino, with the transformation rule
(\ref{tra}), then construct the action as follows
\be
S_0=\int d^{10}x \ L(A,\eta_{MN})
\rightarrow S_\theta=\int d^{10}x \ \sqrt{\cal G}\ 
L(A, {\cal G}_{MN}).
\ee 
The resulting action is invariant under (\ref{tran}) and
(\ref{tra}).
The first two terms of the expansion of the action in $\theta$ read
\be
S_\theta=S_0+{{1}\over{2}}\int d^{10}x \ T^{MN}(A,\eta)\ \bar\theta\Gamma_M
\partial_N \theta+\dots,
\ee
where $T^{MN}$ is the energy momentum tensor of the matter
fields in a Minkowski background. Notice that at this order 
it is sufficient to keep just the first term 
in the transformation of $\theta$, $\delta \theta=\eta$. 
For further details on non-linear supersymmetry, see \cite{sw}. 
\section{Linear N=1 supergravity coupled to \\ non-linear
supersymmetry}

The closed sector of the $USp(32)$ string is described
by an $N=1$ supergravity multiplet which at the sphere
level is supersymmetric. The Lagrangian in the Type I string frame,
obtained by starting with the Einstein frame Lagrangian \cite{cm} and
performing field redefinitions, is given by
\ba
&\!\!\!&L_{SUGRA} \!=\! \sqrt{-g}\Big\{e^{-2\phi}
\Big[-{{R}\over{2}}+2(\partial\phi)^2 \!-\! {{1}\over{2}}
\bar\Psi_M\Gamma^{MNP}D_N\Psi_P
\!+\! 4\bar\lambda\Gamma^MD_M\lambda-\nonumber\\
&\!\!-\!\!&2\sqrt{2}\bar\lambda\Gamma^{MN}D_M\Psi_N \!-\!
2\sqrt{2}\bar\Psi_M\Gamma^N\Gamma^M\lambda\partial_N\phi
\!-\! \bar \Psi_M\Gamma^M\Psi_N\partial^N \phi \Big]
\!-\! {{3}\over{8}} H_{MNP} H^{MNP} \nonumber\\
&\!-\!&e^{-\phi}{{1}\over{16}}
H^{PQR}\Big[\bar\Psi_M\Gamma^{[M}\Gamma_{PQR}\Gamma^{N]}
\Psi_N \!-\! 2\sqrt{2}\bar\Psi_M(\Gamma^{M}{}_{PQR} \!+\!3
\eta^{M}_{P}\Gamma_{QR})\lambda\nonumber\\
&\!-\!&4\bar\lambda \Gamma_{PQR}\lambda\Big]\Big\} \ . \label{sugra}
\ea
It is invariant under the local supersymmetry transformations
\ba
\delta e^A_M &=&{{1}\over{2}}\bar\eta\Gamma^A \Psi_M \ , \ 
\delta \phi = {{1}\over{\sqrt{2}}}\bar\eta\lambda \ , \nonumber \\
\delta B_{MN} &=-&{{1}\over{2}} e^{-\phi}
\bar\eta \ ( \Gamma_M\Psi_N- \Gamma_N \Psi_M
-\sqrt{2} \Gamma_{MN}\lambda) \ , \nonumber \\
\delta \lambda &=& {{1}\over{2\sqrt{2}}}
\Gamma^M\partial_M \phi \ \eta-{{1}\over{8\sqrt{2}}}e^{\phi}
\Gamma^{MNP}H_{MNP} \ \eta \ , \nonumber \\
\delta \Psi_M &=& D_M \eta+{{1}\over{16}}e^\phi \Gamma^{NPQ}
\Gamma_M H_{NPQ} \ \eta \ . \label{linear}
\ea
These transformation rules correspond to the truncation of
the type IIB transformations \cite{sch,brjo} under 
$\Omega'= \Omega(-1)^{F_{spacetime}}$. Notice that
the usual truncation under $\Omega$, which gives rise to the $SO(32)$
Type I string, leads to an equivalent Lagrangian
and transformation rules with the redefinition 
$H\rightarrow -H$. Indeed, consider a Weyl fermion of Type IIB
(gravitino or dilatino),
$\lambda = (\lambda_1, \lambda_2)$, where $\lambda_1, \lambda_2$ are 
Majorana-Weyl fermions. The invariant combination under $\Omega$ is
$\lambda_1+ \lambda_2$, while the one invariant under $\Omega'$ is
$\lambda_1- \lambda_2$. Due to this, the truncation of Type IIB under 
$\Omega'$ gives a opposite sign in the last two lines of (\ref{sugra}) as 
compared to the truncation under $\Omega$. This change of sign is formally
equivalent to the redefinition $H\rightarrow -H$.
Moreover, in passing from the Einstein frame to the Type I frame, in
addition to Weyl rescaling, we also redefined the gravitino field in
order to keep the usual supersymmetry transformation, the first eq. in 
(\ref{linear}). 
 
The open sector of the $USp(32)$ string
has Yang-Mills gauge bosons, one singlet fermion $\theta$
and one fermion multiplet, $\chi$ in the antisymmetric and irreducible
representation of $USp(32)$. At the disk level, we have 
the kinetic terms of these fields, 
 their interaction with the closed sector and the dilaton tadpole
 which receives two equal contributions from the disk and the
 projective plane. When evaluating string three point amplitudes, 
 the only difference between the supersymmetric and 
 the $USp(32)$ calculations at the disk level is with respect
 to the Chan-Paton matrices. This implies an important difference,
 the $\chi\Psi F$ interaction is absent. The $\theta$ fermion has
 no analog in the supersymmetric case so we shall determine its
 contribution to the Lagrangian by consistency.
 At the disk level, the lagrangian reads
 \be
 L_{disk}\!=\!\sqrt{-g}
 \Big\{ e^{-\phi} [-{{1}\over{4}}tr F^2 \!-\! \Lambda \!-\!
{{1}\over{2}} tr \bar \chi
 \Gamma^M D_M(A,\omega)\chi ] \!+\! {{1}\over{16}} tr \bar \chi
\Gamma^{MNP}\chi H_{MNP} \Big\} +L_{\theta} \ . \label{disk}
 \ee
 The existence and the numerical coefficient of the $\chi \chi H$ term
 is found by comparing disk string computations in the $USp(32)$ Type I string
 with the similar ones in the $SO(32)$ Type I string. The vertex
 operator of $H$ is the same in both cases and the one of $\chi$ differs
 only in the Chan-Paton factor, taken into account by the different
 numerical coefficient of the corresponding trace. The precise value of this
 coupling \cite{cm2} will allow a nontrivial string check of our 
 results later on.
    
 The coupling of the non-supersymmetric matter
 to the gravitino implies necessarily that
 there is a non-linear supersymmetry in the matter sector.
 We shall determine $L_\theta$, up to terms bilinear in fermions,
 by the requirement that the complete Lagragian be invariant under
 the above linear transformation supplemented with
 \ba
 \delta \theta &=&\eta \ , \nonumber \\
 \delta A &=& L_v A,\ \delta \chi=L_v\chi \ , \label{volkov} 
 \ea
 where $v$ is defined in (\ref{lie}).
 The variation of $L_{disk}$,  neglecting terms in four or more fermions, 
 reads
 \ba
 \delta L_{disk}&=&\sqrt{-g}\ \tau^{MN}(\phi,F) \
 \bar\eta \Gamma_{M}\Psi_N
 -{{1}\over{\sqrt{2}}}\bar\eta\lambda \ L_{disk}(\phi,F)\\
 &-&{{1}\over{4}}\sqrt{-g} \ 
 \bar\theta\Gamma^M \eta \ tr [F_{MN}D_P(e^{-\phi} F^{PN})]
 +\delta L_\theta \ . \label{var}
 \ea
 The first term comes from the variation of the metric, the second
 from the
 variation of the dilaton, the third from the variation of $A$ and
 finally
 in the fourth we have to take account only 
 of the variation of the fermions, $\theta, \ \Psi_M$ and 
 $\lambda$.
 The tensor $\tau^{MN}$ is the bosonic contribution to the
 energy momentum tensor due to
 $L_{disk}$
 \be
 \tau^{MN}={{1}\over{2}}g^{MN}{{L_{disk}}\over{\sqrt{-g}}}
 +e^{-\phi}{{1}\over{2}}tr(F^{MP}F^{N}{}_{P}) \ .
 \ee
 It verifies the useful identity
 \be
 D_M\tau^{MN}=-{{1}\over{2}}\partial^{N}\phi{{L_{disk}}\over{\sqrt{-g}}}
 +{{1}\over{2}}tr[D_M (e^{-\phi}F^{M}{}_P)F^{NP}] \ .
 \ee
The cancellation of the first term
in (\ref{var}) requires the term $-\bar \theta
\tau^{MN}\Gamma_M\Psi_N$. The variation of $\Psi_M$ 
forces to include a new term proportional to
$\bar\theta\tau^{MN}\Gamma_M D_N\theta$. The cancellation
of the second term in (\ref{var}) is possible
if ${{1}\over{\sqrt{2}}}L_{disk}(\phi,F)
\bar \theta \lambda$ is present in $L_\theta$, it induces a new
variation wich can be cancelled by a $\bar \theta \theta$ term.
In summary, $L_\theta$ has to have the following form
\be
L_\theta=\sqrt{-g}\Big\{-\bar \theta\tau^{MN}\Gamma_M\Psi_N
+{{1}\over{2}}\bar\theta\tau^{MN}\Gamma_MD_N\theta+
{{1}\over{2}}\bar\theta {\cal M}\theta\Big\}
+{{1}\over{\sqrt{2}}}L_{disk}(\phi,F)\ 
\bar \theta \lambda.\label{lt}
\ee
The coefficient in front of the kinetic term of $\theta$
was determined by the requirement of the cancellation of the
terms in $\eta D\theta$, which come from the variation of $\theta$
in the kinetic term as well as from the variation of $\Psi_M$.
The other term $\bar \theta (D_{M}\tau^{MN}) \Gamma_N\eta$
which has the same origin, allows to cancel the last term 
in (\ref{var}) as well as the term in $\partial\phi$ in the
variation of $\lambda$ in (\ref{lt}).
 The matrix $\cal M$ can be readily determined to be given by
\be
{\cal M} = {{1}\over{16}} \Big\{\Gamma^{PQR}
 H_{PQR}\left(\Lambda \!-\! {{1}\over{4}}tr(F^2)\right) 
 \!+\! 3tr(F^{N_3P}F^{N_4}{}_{P})H_{N_1N_2N_3} \Gamma^{N_1N_2}{}_{N_4} \Big\} \ .
\ee
Notice that due to symmetry and chirality reasons 
the only quadratic term in $\theta$ is given by
$a_{MNP}\bar \theta\Gamma^{MNP} \theta$.
Therefore quadratic terms in $\theta$ can only cancel variations
of the form $\bar\eta\Gamma^{MNP}\theta$
but not of the form $\bar \eta\Gamma^M\theta$
or $\bar\eta\Gamma^{M_1\dots M_5}\theta$.
The cancellation of terms in $\bar \eta \Gamma^M\partial_M\phi \
\theta$ between the variation of the kinetic term of 
theta and the term in
$\bar \theta\lambda$ was crucial for the supersymmetric
completion to work.

There is a nice geometrical construction of the 
$\theta$ dependent Lagrangian determined above which is a
generalisation of the setting presented in section II.
Consider the modified moving basis, metric and dilaton defined by
\ba
E^{a}&\!\!=\!\!&e^a+{{1}\over{4}}\bar \theta\Gamma^a D_M\theta dx^M
-{{1}\over{2}}\bar\theta\Gamma^a\Psi_M dx^M+{{1}\over{2}}
\bar \theta S^{a}{}_{b}\theta e^b \ , \ G_{MN}=E^a_M
E^b_N \eta_{ab} \ , \nonumber \\
\hat \phi &\!\!=\!\!& \phi-{{1}\over{\sqrt{2}}}\bar \theta\lambda -
{{1}\over{32}} e^{\phi} \bar \theta\Gamma^{PQR}\theta H_{PQR} \ , \label{mod1}
\ea
with
\be
S^{ab}=-{{1}\over{32}}e^{\phi} [
\Gamma^{PQR}H_{PQR}\eta^{ab} -
6 H_{PQ}{}^{R}\Gamma^{PQ(b}e^{a)}_{R} ] \ . \label{mod2}
\ee
Then, by using (\ref{linear}), it is easily verified that, up to terms 
in (fermi)$^3$,  $E^{a}$ and $\hat \phi$ transform under local 
supersymmetry as
\be
\delta \hat \phi=L_v {\hat \phi} \ , \ \ \delta E^a=L_v
E^a+\Lambda^a{}_{b}E^b \ ,
\ee
where $\Lambda_{ab}$ is antisymmetric and so represents a local
Lorentz rotation. The $E^a$ and $\hat \phi$ transform as do
a moving basis and a scalar under a reparametrisation and a
local Lorentz rotation. In order, to construct
a locally supersymmetric action it is sufficient to replace the 
moving basis by $E^a$ and the dilaton by $\hat \phi$.
Then the action
\ba
{\cal L}_{disk} &=& -\sqrt{-G} \
e^{-\hat \phi} \Big[{{1}\over{4}}G^{MP}G^{NQ} \ tr (F_{MN}F_{PQ}) 
 + \Lambda\Big] \nonumber \\
&+& \sqrt{-g} \Biggl\{ - {{1}\over{2}} tr \bar \chi
 \Gamma^M D_M(A,\omega)\chi + {{1}\over{16}} tr \bar \chi
\Gamma^{MNP} \chi H_{MNP} \Biggr\}
\label{dbi}
\ea
reproduces, when expanded up to quadratic terms in $\theta$,
the complete action $L_{disk}+L_{\theta}$ found above. 
The modified moving basis and dilaton suggest a superspace formulation,
which would allow a comparison with the approach followed in
\cite{abkz,brjo}.

\section{Inclusion of the Chern-Simons terms}

The cancellation of gauge anomalies requires a modification
of the Bianchi identity of the three-form
\be
dH=-{{1}\over{3}} \beta \ tr (F^2) \ . \label{bi}
\ee
More precisely, one-loop anomaly cancellation will fix $\beta=1$,
however for transparency we keep $\beta$ unfixed for the moment. 
The modified relation (\ref{bi}) implies that
\be
H=dB-{{1}\over{3}}\beta \ \omega_3 \ , \label{om}
\ee
and that the variation of $B$ under a gauge transformation
is given by $\delta_{gauge}B=\beta\omega_2^1 /3 $, where $\delta_{gauge}
\omega_3 = d \omega_2^1$.

The lagrangian (\ref{sugra}) with the modified Bianchi
identity is no longer invariant under the transformations
of pure N=1 supergravity. Indeed, the Bianchi identity with 
$\beta=0$ was used when integrating by parts
the variation of the $\Psi D\Psi$, $\lambda D\Psi$,
 $\lambda D \lambda$, $\lambda\Psi H$  
and the $\Psi\Psi H$ terms. A non-zero $\beta$ gives the following 
contribution to the variation of these terms
\be
- \sqrt{-g} \ {{\beta}\over{16}}e^{-\phi}\bar \eta\Gamma^{MNPQR}(
\Psi_R-{{\sqrt{2}}\over{6}}\Gamma_R\lambda) \ tr (F_{MN}F_{PQ}) 
\ . \label{cs1}
\ee
There is a second contribution wich comes from 
the kinetic term of the $H$ field. The latter 
due to (\ref{om}) gets an extra variation originating
from the variation of $\omega_3$ under (\ref{volkov}):
\be
\delta \omega_3=L_v\omega_3=\iota_v tr F^2+d\iota_v\omega_3 \ ,
\ee
where $\iota_v$ is the interior product with respect to the
vector field $v$. 
In order for the variations under a gauge and local supersymmetry
transformations to commute we have to require an extra
variation of $B$:
\be
\delta_\beta B={{1}\over{3}}\beta\iota_v \omega_3 \ , \label{extra}
\ee
that leads to the cancellation of the derivatives of $\eta$
in the extra variation of $H$
\be
\delta_\beta H=-{{1}\over{3}}\beta  \iota_v tr F^2.\label{bh}
\ee 
The variation of $L_{SUGRA}$ due to 
(\ref{bh}) reads
\be
\sqrt{-g} \ {{3}\over{8}} \beta \ \bar \eta\Gamma^M \theta \ 
tr (F_{[MP}F_{QR]}) H^{PQR} \ . \label{kin}
\ee
In order to cancel the term in $\bar\eta\Psi$ 
we have to add to the Lagrangian the piece
\be 
L_\beta^{(1)}=\sqrt{-g}
{{\beta}\over{16}}e^{-\phi}\bar \theta\Gamma^{MNPQR}
\Psi_R \ tr (F_{[MN}F_{PQ]}) \ .
\ee
The variation of $\theta$ in $ L_\beta^{(1)}$ cancels
the first term in (\ref{cs1}). The variation of $\Psi$ induces new terms,
which after integrations by part read
\ba
&\!\!\!& \!\!-\! \sqrt{-g} {{\beta}\over{16}} e^{-\phi} \Biggl\{ 
 D_R {\bar \theta} \ \Gamma^{MNPQR} \ \eta \nonumber \\
&\!+\!& \bar \theta \Gamma^{MNPQR}
(\eta D_R\!-\! \eta \partial_R \phi\!-\!{{1}\over{16}} 
e^\phi \Gamma^{N_1N_2N_3}
\Gamma_R H_{N_1N_2N_3} \eta ) \Biggr\} \ tr (F_{MN}F_{PQ}) \ .
\ea
The first term cancels provided we add to the lagrangian
\be
L_\beta^{(2)}= -\sqrt{-g}{{\beta}\over{32}}e^{-\phi}
\bar \theta\Gamma^{MNPQR}D_R\theta \ tr (F_{MN}F_{PQ}) \ .
\ee
This gives additional variations which are identical to the $D_R$
and $\partial_R\phi$ terms with a factor $-1/2$.
The term in $D_R$ is zero because $tr F^2$ is closed.
The term in $\partial_R\phi$ cannot be cancelled by a quadratic
term in $\theta$. However, it can be cancelled by
a term of the form $\theta\lambda$. Such a term must be present
also to cancel the $\eta\lambda$ terms. It is fixed to be
\be
L_\beta^{(3)}= - \sqrt{-g} {{\sqrt{2}\beta}\over{16}} e^{-\phi}  
\ \bar \theta\Gamma^{MNPQ}\lambda \ tr (F_{MN}F_{PQ}) \ .
\ee
Finally, taking into account the
variation of $L_{SUGRA}$
due to the three-form kinetic term (\ref{kin}),
a supersymmetric completion is now possible if
\ba
&& \Big[ \Gamma^{MNPQR} (\eta {{1}\over{2}}\partial_R\phi+
{{1}\over{16}}e^\phi \Gamma^{N_1N_2N_3}
\Gamma_R H_{N_1N_2N_3} \eta)-
\sqrt{2}\Gamma^{MNPQ}\delta\lambda\nonumber\\
&-& 6  \Gamma^M \eta H^{NPQ}) \Big]tr (F_{MN}F_{PQ})
\ea
can be cast in the form $a_{M_1M_2M_3} \Gamma^{M_1M_2M_3} \ \eta$. 
If we use the relation
\ba
&&\!\! \Bigg\{ \!\! \Gamma^{MNPQR}\Gamma^{N_1N_2N_3}\Gamma_R
\!+\! 2\Gamma^{MNPQ}
\Gamma^{N_1N_2N_3} \!-\!
2\Gamma^{MNPQN_1N_2N_3} \nonumber\\ 
&\!\!\!\!-\!\!\!\!&\!\!
72 g^{MN_1} g^{NM_2}\Gamma^{PQN_3} \!+\! 96
g^{N_1Q} g^{N_2P} g^{N_3N}\Gamma^{M} \Bigg\} H_{N_1N_2N_3}
 tr (F_{MN}F_{PQ}) = 0 \ ,
\ea
it can be readily checked that this is indeed the case.  Finally,
we have to add to the lagrangian the contribution
\be
L_\beta^{(4)}={{1}\over{2}} \sqrt{-g} \ \bar \theta \ {\cal N} \ \theta
\ ,
\ee
with
\ba
{\cal N}&=&-{{\beta}\over{8}} \Biggl\{ 
\Big[{{1}\over{16}}  \Gamma^{MNPQN_1N_2N_3}
+{{9}\over{4}}g^{MN_1}g^{NN_2}\Gamma^{PQN_3}\Big]
H_{N_1N_2N_3}\nonumber\\
&+& e^{-\phi} \Gamma^{NPQ}\partial^{M}\phi \Biggr\}
\ tr (F_{MN}F_{PQ}) \ . 
\ea  
The complete contribution from the Chern-Simons term to the lagrangian
is finally obtained by setting $\beta=1$ and summing up the previous contributions
\be
L_{CS} = L_\beta^{(1)} + L_\beta^{(2)} + L_\beta^{(3)} + L_\beta^{(4)} \
. \label{cs2}
\ee
Collecting all the different contributions (\ref{sugra}), (\ref{dbi}),
(\ref{cs2}) together, we find the complete lagrangian bilinear in the fermion fields
\ba
&&L\!\!=\!\! L_{SUGRA} \!+\! \sqrt{-g} \Biggl\{
\!-\!e^{-\phi} [({{1}\over{4}}tr F^2 \!+\! \Lambda) (1\!+
\!{1 \over \sqrt{2}} \bar
\theta \lambda) +{{1}\over{2}} tr \bar \chi
 \Gamma^M D_M(A,\omega)\chi ] \nonumber \\
&+& {{1}\over{16}} tr \bar \chi\Gamma^{MNP}\chi H_{MNP}  
 - \bar \theta\tau^{MN}\Gamma_M\Psi_N
+{{1}\over{2}}\bar\theta\tau^{MN}\Gamma_M D_N \theta \nonumber \\
&&+{{1}\over{32}}\bar\theta \Big\{\Gamma^{PQR}
 H_{PQR}\left(\Lambda-{{1}\over{4}}tr(F^2)\right) 
 + 3 \ tr(F^{N_3P}F^{N_4}{}_{P})H_{N_1N_2N_3}
 \Gamma^{N_1N_2}{}_{N_4} \Big\}\theta\nonumber\\
&& -{1 \over 16} e^{-\phi} {\bar \theta} \Gamma^{MNPQR} (-\Psi_R + {1
\over 2} D_R \theta + {\sqrt{2} \over 6} \Gamma_R \lambda ) \ 
tr(F_{MN}F_{PQ}) \nonumber \\
&&- {1 \over 16} {\bar \theta}  \bigl[ 
( {{1}\over{16}}  \Gamma^{MNPQN_1N_2N_3}
+{{9}\over{4}}g^{MN_1}g^{NN_2}\Gamma^{PQN_3}) H_{N_1N_2N_3}
+ e^{-\phi} \Gamma^{NPQ}\partial^{M}\phi \bigr]\theta  \nonumber \\
&&  tr(F_{MN}F_{PQ}) \Biggr\} \ , \label{cs3}
\ea
where
 \be
\tau^{MN}= {{1}\over{2}} e^{-\phi} \{ - 
({1 \over 4} tr F^2 + \Lambda)g^{MN}
+ tr(F^{MP}F^{N}{}_{P}) \} \  \label{cs4}
 \ee
is the bosonic part of the energy-momentum tensor of the disk
contribution to the effective Lagrangian. By expanding
(\ref{i5}) up to quadratic terms in the fermions, it is straightforward 
to check that (\ref{cs3}) indeed agrees with (\ref{i5}).
\section{Discussion}
 
We have argued that nonlinear supersymmetry guarantees
the consistency of the gravitino interactions in
the $USp(32)$ type I string. As mentioned in the Introduction,
the existence of the singlet fermion with the appropriate
chirality as well as the possiblility to find a
completion of the effective lagrangian are serious hints in favor
of nonlinear supersymmetry. Notice that many nontrivial cancellations 
were necessary in order for the supersymmetric completion to be
possible and compatible with the anomaly cancellation mechanism.

Here, we would like to add an additional argument 
for the existence of nonlinear supersymmetry in the $USp(32)$
string. Consider the string disk amplitude with two open string
fermions and one Ramond-Ramond two-form. It gives rise
in the effective low energy lagrangian to terms
of the form $\chi \chi H$ and $\theta \theta H$.
From the string theory calculation, the only difference between
these amplitudes is with respect to the Chan-Paton factors.
The $\chi \chi H$ term in the lagrangian can be deduced easily
from the type I effective action (where it is fixed by
supersymmetry, \cite{cm2}). On the other hand, we deduced the $\theta \theta
H$ term by nonlinear supersymmetry. The contribution of the
relevant terms in the effective Lagrangian (\ref{cs3}) reads
\be
 e^{-\phi}\Big[ \!-\! {{1}\over{2}} tr \bar \chi
 \Gamma^M D_M(A,\omega)\chi \!-\! {{\Lambda}\over{4}}\bar \theta 
 \Gamma^MD_M\theta\Big] \!+\!
 {{1}\over{16}} tr \bar \chi\Gamma^{MNP}\chi H_{MNP}
 \!+\! {{\Lambda}\over{32}}\bar\theta \Gamma^{PQR} \theta
 H_{PQR} , \label{contri}
 \ee
 where we 
 displayed also the kinetic terms in order to compare the
 normalisation of $\theta$ and $\chi$. A nontrivial test of the
 existence of nonlinear supersymmetry
 is to compare the string theory calculation of the amplitudes 
 with our predictions. In view of the above remarks and the 
 numerical coefficients of the terms in (\ref{contri}), we see that the 
 string theory calculation and the supersymmetric completion \`a la
 Volkov-Akulov indeed agree. By using the same arguments, we can also
 predict couplings of the type $\chi \chi F^2 H$ and $\chi D \chi F^2$
 in the Type I $SO(32)$ string. These terms arise in the supersymmetric 
 completion of the higher-derivative action and be computed along the
 lines followed in \cite{pvw}.

\vskip 16pt
\begin{flushleft}
{\large \bf Acknowledgments}
\end{flushleft}

\noindent We are grateful to I. Antoniadis,
A. Sagnotti  and especially C. Angelantonj for very useful
discussions and comments. E.D. would like to thank the Aspen Center
for Physics for hospitality and J.M. is grateful to the Center for
Advanced Mathematical Sciences in Beirut, where part of this work was 
carried out.  


\end{document}